\documentclass[twoside]{article}
\usepackage{fleqn,espcrc2}
\usepackage{latexsym}
\newcommand{\be}{\begin{equation}}\newcommand{\ee}{\end{equation}}
\newcommand{\bea}{\begin{eqnarray}}\newcommand{\eea}{\end{eqnarray}}
\newcommand{\nn}{\nonumber\\}\newcommand{\p}[1]{(\ref{#1})}
\newcommand{\lb}{\label}

\newcommand{\cD}{{\cal D}}
\newcommand{\cDb}{{\bar{\cal D}}}
\newcommand{\cA}{{\cal A}}
\newcommand{\cAb}{{\bar{\cal A}}}
\newcommand{\cW}{{\cal W}}
\newcommand{\cWb}{{\bar{\cal W}}}
\title{Superbranes and Super Born-Infeld Theories from
Nonlinear Realizations}

\author{S. Bellucci \address[LNF]{INFN, Laboratori Nazionali di Frascati,
P.O. Box 13, I-00044 Frascati, Italy}, E. Ivanov
\address[JINR]{Bogoliubov Laboratory of Theoretical Physics,
JINR, 141 980 Dubna, Moscow region, Russia} and S. Krivonos
\addressmark[JINR]
}

\begin{document}
\begin{abstract}

We describe, on a few instructive examples, a systematic way of deducing the
superfield equations of motion of superbranes in the approach of partial
breaking of global supersymmetry (PBGS) from the nonlinear-realizations
formalism. For D-branes these equations simultaneously represent the
appropriate supersymmetric Born-Infeld theories. We also discuss how to
construct an off-shell superfield action for the $N=2, d=4$
Dirac-Born-Infeld theory corresponding to the partial supersymmetry breaking
$N=4 \rightarrow N=2$ in $d=4$.    \vspace{1pc}
 \end{abstract}

\maketitle

\section{INTRODUCTION}
There is a growing interest in applying the general method of nonlinear
realizations \cite{nonl} to systems with partial breaking of global
supersymmetries (PBGS) \cite{BW0}, in particular, to the superbranes as a
typical example of such systems (see, e.g., \cite{HLP}-\cite{BIK5} and
refs. therein). In this approach, the superbranes are described by the
Goldstone superfields with the manifest linearly realized worldvolume
supersymmetry and the nonlinearly realized rest of the original full target
supersymmetry. The main difficulty one meets on this path is lacking of  a
systematic procedure for constructing the PBGS actions, as opposed, e.g.,  to
the case of the total supersymmetry breaking \cite{VA}, where the invariant
actions can be constructed following the standard prescriptions of the
nonlinear realizations method.

As a partial way out, it was proposed in \cite{BIK3} to use the nonlinear
realizations approach to deduce the equations of motion for various types of
superbranes in a manifestly covariant language of Cartan 1-forms. These
equations are obtained as a direct covariantization of the free
equations and irreducibility constraints for the Goldstone superfields.
Miraculously, in many cases, while deriving such equations, there appears no
need to incorporate the Goldstone superfields associated with the automorphism
groups of the given supersymmetry, including the Lorentz group (though
such superfields are in general required by the nonlinear
realizations  formalism). It proves enough to deal with the Goldstone
superfields parametrizing spontaneously broken part of the translations and
supertranslations. This circumstance greatly simplifies computations
and allows one to get the superfield equations describing the
superbranes wordvolume dynamics in a concise form.

One of the aims of the present talk is to exemplify this approach by
the $D=4$ supermembrane and related to it via $T$-duality
``space-filling'' D2-brane \cite{IK1}, as well as ``space-filling''
D3-brane (in $D=4$) \cite{BG2,RT} and D3-branes in $D=6$ and $D=10$
\cite{BIK4}.  These systems  correspond, respectively, to the PBGS patterns
$N=2 \rightarrow N=1$ in  $d=3$ and $N=2 \rightarrow N=1$, $N=4 \rightarrow
N=2$ and $N=8 \rightarrow  N=4$ in $d=4$. In the case of D-branes the relevant
Goldstone superfields  encompass the abelian vector multiplets of unbroken
(worldvolume) supersymmetry,  with the Born-Infeld (BI) dynamics for the gauge
field.  Therefore in these cases the obtained superfield equations can be
equivalently treated as the equations of $N=1$, $N=2$ and $N=4$ supersymmetric
BI theories  with hidden second  supersymmetries.

As another topic, we discuss how to construct the full off-shell action
for the $N=2$ BI theory with hidden $N=4$ supersymmetry in $d=4$, corresponding
to the PBGS option  $N=4 \rightarrow N=2$. While for the $N=2 \rightarrow N=1$
BI theory such an action was known \cite{CF,BG2}, only partial results existed
concerning an analogous $N=2$ BI action \cite{Ket,KT}. We present, following
the recent preprint \cite{BIK5}, a systematic procedure for constructing $N=2$
BI action. It is based on embedding of $N=2$ vector multiplet into an
infinite-dimensional  multiplet of the central-charge extended $N=4$, $d=4$
supersymmetry.

\section{N=1,D=4 SUPERMEMBRANE AND SPACE-FILLING D2-BRANE}
Let us start from the well known systems with partially broken
global supersymmetries \cite{AGIT,IK1}. Our goal is to get the corresponding
superfield equations of motion in terms of the worldvolume superfields
starting from the nonlinear realization of the global supersymmetry group.

The supermembrane in $D=4$ spontaneously breaks half of the $N=1,D=4$
supersymmetry and one translation. Let us split the set of generators of
$N=1\; D=4$ Poincar\'e superalgebra
(in the $d=3$ notation) into the unbroken $\left\{ Q_a, P_{ab} \right\}$ and
broken $\left\{ S_a, Z
\right\}$ ones ($a,b=1,2$). The $d=3$ translation generator $P_{ab} =
P_{ba}$ together
with the generator $Z$ form the $D=4$ translation generator.
The basic anticommutation relations read \footnote{Hereafter, we consider the
spontaneously broken supersymmetry algebras modulo possible extra
central-charge type terms which should be present in the full algebra of the
corresponding Noether currents to evade the no-go theorem of ref.
\cite{Wit} along the lines of ref. \cite{HLP}.}
\be \left\{
Q_{a},Q_{b}\right\}=\left\{ S_{a},S_{b}\right\}=P_{ab},\; \left\{
Q_{a},S_{b}\right\} = \epsilon_{ab}Z \;.\label{susy3d}
\ee

In contrast to the consideration in \cite{IK1,BIK1,BIK2}, here
we prefer to deal with the nonlinear realization of the superalgebra \p{susy3d}
itself, ignoring all  generators of the automorphisms of \p{susy3d} (the
spontaneously broken as well as unbroken ones), including those of
$D=4$ Lorentz group $SO(1,3)$. Thus, we
put all generators  into the coset and associate
the $N=1\,,\, d=3$ superspace coordinates $\left\{
\theta^a, x^{ab} \right\}$ with $Q_a, P_{ab}$. The remaining coset
parameters are
Goldstone superfields, $\psi^a \equiv \psi^a(x,\theta),\;q \equiv
q(x,\theta)$. A coset element $g$ is defined by
\be\label{coset3d}
g=e^{x^{ab}P_{ab}}e^{\theta^{a}Q_{a}}e^{qZ}
  e^{\psi^aS_a} \;.
\ee
As the next step, one constructs the
Cartan 1-forms
\bea
&&g^{-1}d g =  \omega_Q^aQ_a + \omega_P^{ab} P_{ab} + \omega_ZZ +
\omega_S^a S_a , \label{cartan13d} \\
&&
\omega_P^{ab} =dx^{ab}+\frac{1}{4}\theta^{(a}d\theta^{b)} +
  \frac{1}{4}\psi^{(a} d\psi^{b)}  \; ,\nn
&&\omega_Z  =   dq+\psi_{a}d\theta^{a}, \;
\omega_Q^a =  d\theta^{a} ,\; \omega_S^a=d\psi^{a}\;.
\label{cartan3d}
\eea
and define the covariant derivatives
\be\label{cd3d}
{\cal D}_{ab} =  (E^{-1})^{cd}_{ab}\,\partial_{cd} , \;
{\cal D}_a = D_a + \frac{1}{2}\psi^b D_a \psi^c \,{\cal D}_{bc},
\ee
where
\be
D_a=\frac{\partial}{\partial \theta^a}+
\frac{1}{2}\theta^b\partial_{ab}, \;
\left\{ D_a, D_b \right\} =\partial_{ab} \; , \label{flatd3d}
\ee
$$ E_{ab}^{cd}=\frac{1}{2}(\delta_a^c\delta_b^d+\delta_a^d\delta_b^c)+
  \frac{1}{4}(\psi^c\partial_{ab}\psi^d+ \psi^d\partial_{ab}\psi^c) \;.
$$
They obey the following algebra
\bea
&&\left[ {\cal D}_{ab},{\cal D}_{cd} \right] =
-{\cal D}_{ab}\psi^f{\cal D}_{cd}\psi^g \,
           {\cal D}_{fg} \; , \nn
&&\left[ {\cal D}_{ab},{\cal D}_{c} \right] =
{\cal D}_{ab}\psi^f{\cal D}_{c}\psi^g\,
           {\cal D}_{fg} \; , \nn
&&\left\{ {\cal D}_{a},{\cal D}_{b} \right\} ={\cal D}_{ab}+
        {\cal D}_{a}\psi^f{\cal D}_{b}\psi^g\,
           {\cal D}_{fg} \; .  \label{algebra3d}
\eea
Not all of the above Goldstone superfields
$\left\{ q(x,\theta),\psi^a(x,\theta)\right\}$
must be treated as independent.
Indeed,  $\psi_{a}$ appears
inside the form $\omega_Z$  {\it linearly} and so can be covariantly
eliminated by the manifestly covariant constraint (inverse Higgs effect
\cite{invh})
\be
\left. \omega_Z\right|_{d\theta} = 0 \Rightarrow \psi_a={\cal D}_a q \;,
\label{basconstr3d}
\ee
where $|_{d\theta}$ means the ordinary $d\theta$-projection of the form.
Thus the superfield
$q(x,\theta)$ is the only essential Goldstone superfield needed to
present the partial spontaneous breaking $N=1\,,\; D=4
\;\Rightarrow \; N=1\,,\; d=3$ within the coset scheme.

In order to get dynamical equations, we put additional,
manifestly covariant constraints on the
superfield $q(x,\theta)$.  The idea is to
covariantize the ``flat'' equations of motion. Namely, we replace
the flat covariant derivatives in the standard equation of motion for
the bosonic scalar superfield in $d=3$
\be\label{flateom13d}
D^a D_a q=0
\ee
by the covariant ones \p{cd3d}
\be\label{eom13d}
{\cal D}^a {\cal D}_a q=0 \;.
\ee
The equation \p{eom13d} coincides with the equation of motion of the
supermembrane in $D=4$ as it was presented in \cite{IK1}. Thus, we conclude
that, at least in this specific case, additional superfields-parameters of the
extended coset with all the automorphism symmetry generators included are
auxiliary and can be dropped out if we are interested in the equations of
motion only. Actually, it can be shown that \p{eom13d} possesses the hidden
covariance under the full $D=4$ Lorentz group.

As a straightforward application of the idea that the automorphism
symmetries are irrelevant when deducing the equations of motion,
let us consider the case of the ``space-filling''
D2-brane (i.e. having $N=1,\,d=3$ vector multiplet as its
worldvolume one).

The superalgebra we start with is the algebra \p{susy3d}
without the central charge,$ Z=0 $.
The coset element $g$ contains only one Goldstone superfield $\psi^a$
which now must be treated as the essential one, and the covariant derivatives
coincide
with \p{cd3d}. Bearing in mind to end up with the irreducible field
content of $N=1,\,d=3$ vector multiplet, we are led to treat
$\psi^a$ as the corresponding  superfield strength and to find the
appropriate covariantization of the  flat irreducibility constraint and the
equation of motion. In the flat case the $d=3$ vector multiplet is represented
by a $N=1$ spinor superfield strength $\mu_a$ subjected to the Bianchi identity
\cite{bibl}:
\be\label{cc13d}
D^a\mu_a=0 \; \Rightarrow \; \left\{
 \begin{array}{l}
   D^2 \mu_a=-\partial_{ab}\mu^b~,  \nn
   \partial_{ab}D^a\mu^b = 0~ . \nonumber
  \end{array} \right.
\ee
This leaves in $\mu_a$ the first fermionic
(Goldstone) component, together with the
divergenceless vector $F_{ab}\equiv D_a\mu_b|_{\theta=0}$
(i.e., just the gauge field strength). The equation of motion reads
\be\label{flateom23d}
D^2 \mu_a =0 \; .
\ee
In accordance with our approach, we propose the following equations
which should describe the D2-brane:
\be\label{eom23d}
 (a)\;\;{\cal D}^a\psi_a=0~, \quad (b)\; \; {\cal D}^2 \psi_a =0~.
\ee

In order to see which kind of dynamics is encoded in \p{eom23d}, we
considered it
in the bosonic limit. We found that it amounts to the following
equations
for the vector $V_{ab}\equiv {\cal D}_a\psi_b|_{\theta=0}$:
\be\label{boseq13d}
\left( \partial_{ac} +V_a^m V_c^n \partial_{mn}\right) V_b^c =0 \;.
\ee
One can wonder how these nonlinear but polynomial equations can be related
to the nonpolynomial  BI theory which is just the bosonic core of
the superfield D2-brane theory as was explicitly demonstrated in \cite{IK1}.
The trick is to rewrite the antisymmetric and symmetric parts of the equation
\p{boseq13d} as follows:
\bea
&& \partial_{ab} \left( \frac{V^{ab}}{2-V^2} \right)=0 \;,
\label{boseq23d}\\
&& \partial_{ac} \left( \frac{ V_b^c}{2+V^2}\right)+
 \partial_{bc} \left( \frac{ V_a^c}{2+V^2}\right) =0 \;, \label{boseq33d}
\eea
where $V^2\equiv V^{mn}V_{mn}$.
After passing to the ``genuine'' field strength
\be\label{fs3d}
F^{ab}=\frac{2V^{ab}}{2-V^2} \Rightarrow
\partial_{ab}F^{ab}=0\;,
\ee
the equation of motion \p{boseq33d} takes the familiar BI form
\be
\partial_{ac} \left( \frac{ F_b^c}{\sqrt{1+2F^2}}\right)+
 \partial_{bc} \left( \frac{ F_a^c}{\sqrt{1+2F^2}}\right) =0 \;.
\label{boseq43d}
\ee
Thus we have proved that the bosonic part of our system \p{eom23d} indeed
coincides with the BI equations. One may explicitly show that the full
equations \p{eom23d} are equivalent to the worldvolume superfield equation
following from the off-shell D2-brane action given in \cite{IK1}.

\section{SPACE-FILLING D3-BRANE}
As another example we consider the
space-filling D3-brane in $d=4$. This system amounts to the PBGS pattern
$N=2 \rightarrow  N=1$ in $d=4$, with a nonlinear generalization of
$N=1, \, d=4$ vector multiplet as the Goldstone multiplet \cite{BG2,RT}. The
off-shell superfield action for this system and the related equations of
motion are known \cite{BG2}, but the latter have never been derived directly
from the coset approach.

Our starting point is the $N=2,\, d=4$ Poincar\'e superalgebra {\it
without}
central charges:
\be
\left\{ Q_{\alpha}, {\bar Q}_{\dot\alpha} \right\}=2P_{\alpha\dot\alpha}
\;,\;
\left\{ S_{\alpha}, {\bar S}_{\dot\alpha} \right\}=2P_{\alpha\dot\alpha}
\;.\label{susyD43d}
\ee
Assuming the $S_{\alpha}, {\bar S}_{\dot\alpha}$ supersymmetries
to be spontaneously broken, we introduce the Goldstone superfields
$\psi^{\alpha}(x,\theta,\bar\theta), \,
{\bar\psi}^{\dot\alpha}(x,\theta,\bar\theta)$ as the corresponding parameters
in the following coset (we use the same notation as in \cite{BG2})
\be
g=e^{ix^{\alpha\dot\alpha}P_{\alpha\dot\alpha}}
   e^{ i\theta^{\alpha}Q_{\alpha}+
   i{\bar\theta}_{\dot\alpha}{\bar Q}^{\dot\alpha}}
e^{ i\psi^{\alpha}S_{\alpha}+
   i{\bar\psi}_{\dot\alpha}{\bar S}^{\dot\alpha}} \;.
\ee
With the help of the Cartan forms
\bea\label{cf53d}
g^{-1}dg & =& i\omega^{\alpha\dot\alpha}P_{\alpha\dot\alpha}+
  i\omega_Q^{\alpha}Q_{\alpha}+
  i{\bar\omega}_{Q\;\dot\alpha}{\bar Q}^{\dot\alpha}\nn
  &&+i\omega_S^{\alpha}S_{\alpha}+
  i{\bar\omega}_{S\;\dot\alpha}{\bar S}^{\dot\alpha}\;,\\
\omega^{\alpha\dot\alpha} & = & dx^{\alpha\dot\alpha}-i\left(
  \theta^{\alpha}d{\bar\theta}^{\dot\alpha}\right.\nn
 &&+\left.
  {\bar\theta}^{\dot\alpha}d\theta^{\alpha}+
\psi^{\alpha}d{\bar\psi}^{\dot\alpha}+
  {\bar\psi}^{\dot\alpha}d\psi^{\alpha}\right) \;, \nonumber
\eea
$$
\omega_Q^{\alpha}=d\theta^{\alpha},\;
{\bar\omega}_Q^{\dot\alpha}=d{\bar\theta}^{\dot\alpha},\;
\omega_S^{\alpha}=d\psi^{\alpha},\;
{\bar\omega}_S^{\dot\alpha}=d{\bar\psi}^{\dot\alpha},
$$
one can define the covariant derivatives
\bea\label{cd43d}
&&{\cal D}_{\alpha}=  D_{\alpha}-i\left(
{\bar\psi}^{\dot\beta}D_{\alpha}\psi^{\beta} +
  \psi^{\beta}D_{\alpha}{\bar\psi}^{\dot\beta}\right) {\cal
   D}_{\beta\dot\beta},  \nn
&&{\cal D}_{\alpha\dot\alpha}=
 \left(
E^{-1}\right)_{\alpha\dot\alpha}^{\beta\dot\beta}\partial_{\beta\dot\beta}
  \;,
\eea
where
\be
E_{\alpha\dot\alpha}^{\beta\dot\beta}= \delta_{\alpha}^{\beta}
 \delta_{\dot\alpha}^{\dot\beta}
-i\psi^{\beta}\partial_{\alpha\dot\alpha}{\bar\psi}^{\dot\beta}-
i{\bar\psi}^{\dot\beta}\partial_{\alpha\dot\alpha}\psi^{\beta}\;,
\ee
and the flat covariant derivatives are defined as follows
\be\label{flat4d3d}
D_{\alpha}=\frac{\partial}{\partial\theta^{\alpha}} -
i{\bar\theta}^{\dot\alpha}\partial_{\alpha\dot\alpha}~, \;
{\bar D}_{\dot\alpha}=
 -\frac{\partial}{\partial{\bar\theta}^{\dot\alpha}} +
i{\theta}^{\alpha}\partial_{\alpha\dot\alpha}\;.
\ee
Now we are ready to write the covariant version of the constraints
on $\psi^{\alpha},\,{\bar\psi}^{\dot\alpha}$ which define the superbrane
generalization of $N=1,\,d=4$ vector multiplet, together with the covariant
equations of motion for this system.

As is well-known \cite{bw}, the $N=1,\,d=4$ vector multiplet is
described by
a chiral $N=1$ field strength $W_\alpha \,$,
\be\label{chir53d}
{\overline D}_{\dot\alpha}W_{\alpha}=0~, \quad
D_{\alpha}{\overline W}_{\dot\alpha}=0~, \ee
which satisfies the irreducibility constraint (Bianchi identity)
\be\label{constr53d}
D^{\alpha}W_{\alpha}+{\overline D}_{\dot\alpha}{\overline
W}^{\dot\alpha}=0
\;.
\ee
The free equations of motion for the vector multiplet read
\be\label{eom53d}
D^{\alpha}W_{\alpha}-{\overline D}_{\dot\alpha}{\overline
W}^{\dot\alpha}=0
\;.
\ee

It was shown in \cite{BG2} that the chirality constraints \p{chir53d} can
be directly covariantized
\be\label{chir5e3d}
{\overline{\cal D}}_{\dot\alpha}\psi_{\alpha}=0~, \quad {\cal
D}_{\alpha}\bar\psi_{\dot\alpha}=0~.
\ee
These conditions are compatible with the algebra of the
covariant  derivatives \p{cd43d}. This algebra, with the constraints
\p{chir5e3d} taken into
account, reads \cite{BG2}
\bea && \{{\cal D}_\alpha, \,{\cal D}_\beta \} =
\{{\overline{\cal D}}_{\dot\alpha}, \, {\overline{\cal D}}_{\dot\beta} \} =
0~, \nn && \{{\cal D}_\alpha, \,{\overline{\cal D}}_{\dot\beta} \} = 2i\,{\cal
D}_{\alpha \dot\beta} - 2i\,({\cal D}_{\alpha}\psi^\gamma {\overline{\cal
D}}_{\dot\beta}  {\bar\psi}^{\dot\gamma})\, {\cal D}_{\gamma\dot\gamma}~, \nn
&&\{{\cal D}_\alpha, \,{\cal D}_{\gamma\dot\gamma}\} = -2i\,( {\cal
D}_{\alpha}\psi^\beta {\cal D}_{\gamma\dot\gamma} {\bar\psi}^{\dot\beta})\,
{\cal D}_{\beta\dot\beta}~. \label{cd4alg3d}
\eea
The first two relations in \p{cd4alg3d} guarantee the consistency of the above
nonlinear version of $N=1, \;d=4$ chirality. They also imply, like in the flat
case,
\be
({\cal D})^3 = ({\overline{\cal D}})^3 = 0~. \label{cubvan3d}
\ee

The second flat irreducibility constraint, eq.\p{constr53d},
is not so simple to covariantize. The straightforward
generalization of \p{constr53d},
\be\label{constr5f3d}
{\cal D}^{\alpha}\psi_{\alpha}+
 {\overline{\cal D}}_{\dot\alpha}{\overline \psi}^{\dot\alpha}=0~,
\ee
is contradictory. Let us apply the square $({\cal D})^2$ to
the  left-hand side of \p{constr5f3d}. When hitting the first term in the sum,
it yields zero in virtue of the property \p{cubvan3d}. However, it
is not zero on the second term. To compensate
for the resulting non-vanishing terms, and thus to achieve compatibility
with the algebra \p{cd4alg3d} and its corollaries \p{cubvan3d}, one should
modify \p{constr5f3d} by some  higher-order corrections
\cite{BG2}.

Let us argue that the constraints \p{constr53d} {\it together} with the
equations of motion \p{eom53d} can be straightforwardly covariantized as
\be\label{constr5h3d}
{\cal D}^{\alpha}\psi_{\alpha}=0\;,\quad  {\overline{\cal
D}}_{\dot\alpha}{\overline \psi}^{\dot\alpha}=0 \;.
\ee

Firstly, we note that no difficulties of the above kind related to the
compatibility with the  algebra \p{cd4alg3d} arise on the shell of eqs.
\p{constr5h3d}. As a consequence of \p{constr5h3d} and the first two relations
in \p{cd4alg3d}  we get
\be
{\cal D}^2\,\psi_{\alpha}=0,\quad {\overline{\cal
D}}^2\,\bar\psi_{\dot\alpha}=0\;. \label{squarenonl3d}
\ee
This set is a nonlinear version of the well-known reality condition and
the equation of motion for the auxiliary field of vector
multiplet. Then, applying, e.g.,
${\cal D}_\alpha$ to the second equation in \p{constr5h3d} and making use of
the chirality condition \p{chir5e3d}, we obtain the nonlinear version of the
equation of motion for photino
\be
{\cal D}_{\alpha \dot\alpha}\bar\psi^{\dot\alpha} - ({\cal
D}_{\alpha}\psi^\gamma {\overline{\cal D}}_{\dot\alpha}
{\bar\psi}^{\dot\gamma})\, {\cal D}_{\gamma\dot\gamma}\bar\psi^{\dot\alpha} =
0~.
\ee
Acting on this equation by one more ${\cal D}_\alpha$ and taking advantage of
the equations \p{squarenonl3d} we obtain the identity $0 = 0$, i.e. no new
dynamical  restrictions on $\psi^\alpha, \bar\psi^{\dot\alpha}$ arise. On the
other hand, acting by ${\overline{\cal D}}_{\dot\alpha}$ produces a nonlinear
generalization of the Maxwell equation. Further applying the covariant
derivatives to this equation does not lead to new consequences.
It can be also explicitly checked,
in a few lowest orders in $\psi^\alpha, \bar\psi^{\dot\alpha}$, that the
higher-order corrections to \p{constr5f3d} found in \cite{BG2} are vanishing
on the shell of eqs. \p{constr5h3d}.

Thus the full set of equations describing the dynamics of the D3-brane
supposedly consists of the generalized chirality constraint \p{chir5e3d}
and the equations \p{constr5h3d}. To prove its equivalence to the $N=1$
superfield description of D3-brane proposed in \cite{BG2}, recall that
the latter is the $N=1$ supersymmetrization \cite{CF} of the $d=4$ BI
action  with one extra nonlinearly realized $N=1$ supersymmetry. So, let us
consider the bosonic part of the proposed set of equations.
Our superfields
$\psi,\bar\psi$ contain the following bosonic components:
\be\label{defcomp53d}
V^{\alpha\beta}\equiv {\cal
D}^{\alpha}\psi^{\beta}|_{\theta=0}~, \; {\bar V}^{{\dot\alpha}\dot\beta}=
\equiv   {\overline{\cal
D}}^{\dot\alpha}{\bar\psi}^{\dot\beta}|_{\theta=0},
\ee
which, owing to \p{constr5h3d}, obey the following simple equations
\bea\label{eom63d}
&&\partial_{\alpha\dot\alpha}V^{\alpha\beta}-V_{\alpha}^{\gamma}
 {\bar V}_{\dot\alpha}^{\dot\gamma}\;\partial_{\gamma\dot\gamma}
  V^{\alpha\beta} =0~,\nn
&&\partial_{\alpha\dot\alpha}{\bar
V}^{{\dot\alpha}\dot\beta}-V_{\alpha}^{\gamma}
 {\bar V}_{\dot\alpha}^{\dot\gamma}\;\partial_{\gamma\dot\gamma}
  {\bar V}^{{\dot\alpha}\dot\beta} =0 \;.
\eea
Like in the D2-brane case, in the equations \p{eom63d} nothing
reminds us of the BI equations. Nevertheless, it is possible to
rewrite these equations in the standard BI form.

After some algebra, one can bring eqs.\p{eom63d} into the following equivalent
form
\bea\label{last13d}
&&\partial_{\beta\dot\alpha}\left(fV_{\alpha}^{\beta}\right) -
  \partial_{\alpha\dot\beta}\left( {\bar f}{\bar
V}_{\dot\alpha}^{\dot\beta}
  \right) =0~, \nn
&&\partial_{\beta\dot\alpha}\left(gV_{\alpha}^{\beta}\right) +
  \partial_{\alpha\dot\beta}\left( {\bar g}{\bar
V}_{\dot\alpha}^{\dot\beta}
  \right) =0 \;,
\eea
where
\be\label{sol13d}
f=\frac{ {\bar V}{}^2-2}{1-\frac{1}{4}V^2{\bar V}{}^2}~, \quad
g=\frac{ {\bar V}{}^2+2}{1-\frac{1}{4}V^2{\bar V}{}^2}\; .
\ee
After introducing the ``genuine'' field strengths
\be\label{fc53d}
F_\alpha^\beta\equiv \frac{1}{2\sqrt{2}}\, f\,V_\alpha^\beta~, \quad
{\bar F}_{\dot\alpha}^{\dot\beta}\equiv \frac{1}{2\sqrt{2}}\,
  {\bar f}\,{\bar V}_{\dot\alpha}^{\dot\beta}~,
\ee
first of eqs. \p{last13d} is recognized as the Bianchi
identity
\be
\partial_{\beta\dot\alpha}F_{\alpha}^{\beta} -
  \partial_{\alpha\dot\beta}{\bar F}_{\dot\alpha}^{\dot\beta}   =0 \;,
\ee
while the second one acquires the familiar form of the BI equation
\be\label{last23d} \partial_{\beta\dot\alpha}\left(
\frac{1+A}{B}
F_{\alpha}^{\beta}\right)
 + \partial_{\alpha\dot\beta}\left(
\frac{1-A}{B}
{\bar F}_{\dot\alpha}^{\dot\beta}
  \right) =0,
\ee
where
\bea
&&A=F^2-{\bar F}{}^2 \; , \nn
&&B=\sqrt{(F^2-{\bar F}{}^2)^2-2(F^2+{\bar
F}{}^2)+1} \; .
\eea
Thus, in this new basis the action for our bosonic system is
the BI action:
\be
S=\int d^4x \sqrt{(F^2-{\bar F}{}^2)^2-2(F^2+{\bar F}{}^2)+1} \;.
\ee

Now the equivalence of the system \p{constr5h3d} to the equations corresponding
to the action of ref. \cite{BG2} can be established like in the D2-brane
case.

Note that at the full superfield level the redefinition \p{fc53d} should
correspond to passing from the Goldstone fermions  $\psi_\alpha$,
$\bar\psi_{\dot\alpha}$  which have the simple transformation properties
in the nonlinear realization of $N=1, \,d=4$ supersymmetry but obey the
nonlinear irreducibility constraints, to the ordinary Maxwell superfield
strength $W_\alpha, \,\bar W_{\dot\alpha}$ defined by eqs. \p{chir53d},
\p{constr53d}. The nonlinear action in \cite{BG2} was written just in terms of
this latter  object. The equivalent form \p{constr5h3d} of the equations of
motion and  Bianchi identity is advantageous in that it is manifestly
covariant under  the second (hidden) supersymmetry, being constructed out of
the  covariant objects.

\section{N=2 BI THEORY WITH PARTIALLY BROKEN N=4 SUPERSYMMETRY}
\subsection{Vector Goldstone multiplet for \break $N=4 \rightarrow N=2$}

Now we wish to derive the $N=2$ supersymmetric BI theory as a theory
of the partial breaking of $N=4, d=4$ supersymmetry down to $N=2$
supersymmetry, with the vector $N=2$ multiplet as the Goldstone one.
As follows from the field content of the latter, this kind of $N=2$
BI theory should amount to a static-gauge form of D3-brane in $D=6$.

To apply the nonlinear realizations techniques \cite{BIK3}, we firstly
need to specify $N=4, d=4$ supersymmetry to start with.

In the $N=2$ Maxwell theory, the basic object is a complex scalar
$N=2$ off-shell superfield strength ${\cal W}$ which is chiral and
satisfies one additional Bianchi identity:
\bea
&&(a)\;\bar D_{\dot\alpha i} {\cal W} =0~,\;
D^i_{\alpha} \bar{\cal W} =0~,\nn
&&(b) \;D^{ik}{\cal W} = \bar D^{ik}\bar{\cal  W}~. \lb{oshN2M4d}
\eea
Here,
\be
D_{\alpha}^i = \frac{\partial}{\partial\theta^{\alpha}_i}+
  i{\bar\theta}^{{\dot\alpha}i}\partial_{\alpha\dot\alpha},\;
{\bar D}_{{\dot\alpha}i}=-\frac{\partial}
{\partial{\bar\theta}^{\dot\alpha i}}
  -i\theta^{\alpha}_i\partial_{\alpha\dot\alpha},
\label{semicd4d}
\ee
\be
 D^{ij} \equiv  D^{\alpha\, i}D_{\alpha}^j~, \;
\bar D^{ij} \equiv  \bar D^{i}_{\dot\alpha}D^{\dot\alpha \,j}~,
(i,j=1,2).\lb{defDD4d}
\ee
The superfield equation of motion for ${\cal W}$ reads
\be
D^{ik}{\cal W} + \bar D^{ik}\bar{\cal W} = 0~, \lb{eqm14d}
\ee
and, together with (\ref{oshN2M4d}$b$), amounts to
\be
D^{ik}{\cal W}  = \bar D^{ik}\bar{\cal W} = 0~. \lb{eqmN2M4d}
\ee

In order to incorporate an appropriate generalization of
${\cal W}$ into the nonlinear realization scheme as the
Goldstone superfield, we need to have
the proper bosonic generator in the algebra. The following central
extension of $N=4, d=4$ Poincar\'e
superalgebra suits this purpose
\bea
&&\left\{ Q_{\alpha}^i, {\bar Q}_{\dot{\alpha}j}
        \right\}=
\left\{ S_{\alpha}^i, {\bar S}_{\dot{\alpha}j}
        \right\}=2\delta^i_jP_{\alpha\dot{\alpha}}, \nn
&&\left\{ Q_{\alpha}^i,  S_{\beta}^j
   \right\}=2\varepsilon^{ij}\varepsilon_{\alpha\beta}Z ,\nn
&&\left\{ {\bar Q}_{\dot{\alpha} i}, {\bar Q}_{\dot{\beta}j}
        \right\}=-2\varepsilon^{ij}
      \varepsilon_{\dot\alpha  \dot\beta}{\bar Z} ,
 \label{n4sa4d}
\eea
with all other (anti)commutators vanishing. Note an important feature
that the complex central charge $Z$ appears in the crossing
anticommutator, while the generators $(Q, \bar Q)$ and $(S, \bar S)$
on their own form two $N=2$ superalgebras without central charges.
The full internal symmetry automorphism
group of \p{n4sa4d} (commuting with $P_{\alpha\dot\alpha}$ and $Z$)
is $SO(5)\sim Sp(2)$. Besides the manifest $R$-symmetry
group $U(2)_R = SU(2)_{R}\times U(1)_R$
acting as uniform rotations of the doublet indices of all spinor
generators and the opposite phase transformations of the $S$- and
$Q$-generators, it also includes the 6-parameter quotient
$SO(5)/U(2)_{R}$ transformations
which properly rotate the generators $Q$ and $S$ through each other. The
superalgebra \p{n4sa4d} is a $d=4$ form of
$N=(2,0)$ (or $N=(0,2)$) Poincar\'e superalgebra in $D=6$.

Let us now split the set of generators of the $N=4$ superalgebra \p{n4sa4d}
into the unbroken $\left\{ Q_{\alpha}^i,{\bar Q}_{\dot\alpha j},
P_{\alpha\dot\alpha} \right\}$ and broken $\left\{ S_{\alpha}^i,
{\bar S}_{\dot\alpha j}, Z,{\bar Z} \right\}$ parts and define a coset element
$g$ as:
\bea\label{vectorcoset4d}
g= \mbox{exp}\,i\left(-x^{\alpha\dot{\alpha}}P_{\alpha\dot{\alpha}} +
\theta^{\alpha}_i Q^i_{\alpha}+{\bar\theta}^i_{\dot\alpha}
    {\bar Q}_i^{\dot\alpha}\right)\nn
  \mbox{exp}\,i\left(\psi^{\alpha}_i
S^i_{\alpha}+{\bar\psi}^i_{\dot\alpha}     {\bar S}_i^{\dot\alpha}\right)
   \mbox{exp}\,i\left(WZ+{\bar W}{\bar Z}\right) \;.
\eea
Acting on \p{vectorcoset4d} from the left by various elements
of the supergroup corresponding to \p{n4sa4d}, one can find the
transformation properties of the coset coordinates.

For the unbroken supersymmetry $\left( g_0=\mbox{exp}\,i
\left( -a^{\alpha\dot\alpha}P_{\alpha\dot\alpha}+
\epsilon^{\alpha}_i Q^i_{\alpha}+{\bar\epsilon}^i_{\dot\alpha}
    {\bar Q}_i^{\dot\alpha}\right)\right) $ one has:
\bea\label{unbrokentr4d}
&&\delta x^{\alpha\dot\alpha}=a^{\alpha\dot\alpha} -i\left(
\epsilon^{\alpha}_i{\bar\theta}^{\dot\alpha i}+
{\bar\epsilon}^{\dot\alpha i}\theta^{\alpha}_i\right) , \nn
&&\delta\theta^{\alpha}_i=\epsilon^{\alpha}_i\; , \quad
\delta{\bar\theta}^i_{\dot\alpha}=\bar\epsilon^i_{\dot\alpha} \;.
\eea

Broken supersymmetry transformations
$\left( g_0=\mbox{exp}\,i
\left(
\eta^{\alpha}_i S^i_{\alpha}+{\bar\eta}^i_{\dot\alpha}
    {\bar S}_i^{\dot\alpha}\right)\right) $ are as follows:
\bea\label{brokentr4d}
&&\delta x^{\alpha\dot\alpha}= -i\left(
\eta^{\alpha}_i{\bar\psi}^{\dot\alpha i}+
{\bar\eta}^{\dot\alpha i}\psi^{\alpha}_i\right) \;, \nn
&&\delta\psi^{\alpha}_i=\eta^{\alpha}_i\; , \quad
\delta{\bar\psi}^i_{\dot\alpha}=\bar\eta^i_{\dot\alpha} \;, \nn
&&\delta W = -2i\eta^{\alpha}_i\theta^i_{\alpha} \:, \quad
\delta{\bar W}=-2i{\bar\eta}^i_{\dot\alpha}{\bar\theta}_i^{\dot\alpha}\;.
\eea

Finally, the
broken $Z,{\bar Z}$-translations $\left( g_0=\mbox{exp}\,i
\left( cZ+{\bar c}{\bar Z}\right)\right) $ read
\be\label{brokenZ4d}
\delta W = c \:, \quad
\delta{\bar W}= {\bar c}\;.
\ee

The next standard step is to define
the left-invariant Cartan 1-forms:
\bea\label{vectorcf4d}
\omega_P^{\alpha\dot\alpha} &=& dx^{\alpha\dot\alpha}-
  i\left( d{\bar\theta}^{{\dot\alpha}i} \theta_i^{\alpha} +
  d\theta_i^{\alpha}{\bar\theta}^{{\dot\alpha}i}\right) \nn
&& -i
  \left(d{\bar\psi}^{{\dot\alpha}i} \psi_i^{\alpha}+
  d\psi_i^{\alpha}{\bar\psi}^{{\dot\alpha}i}\right)~, \nn
\omega_{Q\; i}^{\alpha} &=& d\theta^\alpha_i \; , \quad
   \omega_{S\; i}^{\alpha} = d\psi^\alpha_i \; ,  \nn
\omega_Z &=& dW-2i d\theta^\alpha_i  \psi^i_\alpha \; .
\eea
The covariant derivatives of some scalar $N=2$ superfield $\Phi $
are defined by expanding
the differential $d\Phi$ over the covariant differentials of
the $N=2$ superspace coordinates
\bea
&& d\Phi \equiv {\omega}_P^{\alpha\dot\alpha}
\nabla_{\alpha\dot\alpha} \Phi+
 d\theta^\alpha_i \cD_\alpha^i \Phi +
   d\bar\theta_{\dot\alpha}^i{\cDb}^{\dot\alpha}_i \Phi \;\;\Rightarrow \nn
&& \nabla_{\alpha\dot\alpha}=
  \left( E^{-1} \right)_{\alpha\dot\alpha}^{\beta\dot\beta}
  \partial_{\beta\dot\beta}\; , \nn
&& E_{\alpha\dot\alpha}^{\beta\dot\beta}\equiv
\delta_{\alpha}^{\beta}\delta_{\dot\alpha}^{\dot\beta} +i
 \psi_i^{\beta}\partial_{\alpha\dot\alpha}{\bar\psi}^{\dot\beta i}
 +i {\bar\psi}^{\dot\beta i}\partial_{\alpha\dot\alpha}
 \psi^\beta_i \;, \nonumber
\eea
\be
\cD_{\alpha}^i = D_\alpha^i +i \left(
\psi^\beta_j D_{\alpha}^i   {\bar\psi}^{\dot\beta j}+{\bar\psi}^{\dot\beta
j}D_{\alpha}^i       \psi^\beta_j \right)\nabla_{\beta\dot\beta}  ,
\label{fullcd4d}
\ee
where $D^i_\alpha, \bar D_{i\dot\alpha}$ are defined in \p{semicd4d}.

As in previously studied examples,
the Goldstone fermionic superfields
$\psi^i_\alpha$, ${\bar\psi}_{\dot\alpha i}$ can be covariantly
expressed in terms of the central-charge Goldstone superfields
${\cal W}$, $\bar{\cal W}$ by imposing the inverse Higgs
constraints \cite{invh} on the central-charge Cartan 1-forms. In the present
case these constraints are
\be
\omega_Z\vert_{d\theta, d\bar\theta} =
\bar\omega_Z\vert_{d\theta, d\bar\theta} = 0~,  \lb{IH4d}
\ee
where $\vert $ means the covariant projections on the differentials of
the spinor coordinates. These constraints amount to the sought expressions
for the fermionic Goldstone superfields
\be\label{vectorih4d}
\psi^i_\alpha = -\frac{i}{2}\cD_{\alpha}^i W\;,\quad
{\bar\psi}_{\dot\alpha i}=-\frac{i}{2}\cDb_{\dot\alpha i}{\bar W}~,
\ee
and, simultaneously, to the covariantization of the chirality conditions
(\ref{oshN2M4d}$a$)
\be
\cDb_{\dot\alpha i} W =0 \; , \quad \cD_{\alpha}^i {\bar W}=0 \;.
 \label{vectorchir4d}
\ee
Actually, eqs. \p{vectorih4d} are highly nonlinear equations serving to
express $\psi^i_\alpha$, ${\bar\psi}_{\dot\alpha i}$
in terms of $W, \bar W$ with making use of the
definitions \p{fullcd4d}.

It is also straightforward to write the covariant generalization of the
dynamical equation of the $N=2$ abelian vector multiplet \p{oshN2M4d},
\p{eqmN2M4d}
\bea
\cD^{\alpha( i}\cD_{\alpha}^{j)} W= 0\;,\quad
   \cDb_{\dot\alpha}^{(i} \cDb^{\dot\alpha j)}{\bar W} =0 \; .
  \label{vectoreom4d}
\eea

The equations \p{vectorchir4d}, \p{vectoreom4d} with the superfield
Goldstone fermions
eliminated by \p{vectorih4d} constitute a manifestly covariant form of the
superfield equations of motion of $N=2$ Dirac-BI theory with
the second hidden nonlinearly realized $N=2$ supersymmetry. It closes,
together with the manifest $N=2$ supersymmetry,
on the $N=4$ supersymmetry \p{n4sa4d}.

As a first step in proving this statement, let us show that the above system
of equations reduces the component content of $W$ just to that of
the on-shell
$N=2$ vector multiplet. It is convenient to count the number of independent
covariant superfield projections of $W$, $\bar W$.

At the dimensions $(-1)$ and $(-1/2)$ we find $W$, $\bar W$ and
$\psi_{i\alpha} = -\frac{i}{2}\cD_{i\alpha} W\;,\quad
{\bar\psi}_{\dot\alpha}^i = \overline{(\psi_{i\alpha})}
=-\frac{i}{2}\cDb_{\dot\alpha}^i{\bar W}$, with a complex bosonic field and
a doublet of gaugini as the lowest components.

At the dimension $(0)$ we have,
before employing \p{vectorchir4d}, \p{vectoreom4d},
\bea
&& \cD_{\alpha}^i\psi_{\beta}^j=
 \varepsilon^{ij}f_{\alpha\beta} +
i\varepsilon_{\alpha\beta}F^{(ij)} +
F^{(ij)}_{(\alpha\beta)}\; , \nn
&& \cDb_{\dot\alpha i}\psi_{j\alpha} =
\varepsilon_{ij} X_{\alpha\dot\alpha} + X_{(ij)\alpha\dot\alpha},
\label{components114d} \\
&&f_{\alpha\beta} \equiv \epsilon_{\alpha\beta}A + iF_{\alpha\beta},
\; \bar f_{\dot\alpha\dot\beta} =
\epsilon_{\dot\alpha\dot\beta}\bar A - i\bar F_{\dot\alpha\dot\beta},\nn
&& f^\alpha_\beta f_{\alpha\gamma} = \epsilon_{\beta\gamma}\left(A^2 -
{1\over 2}F^2\right)~.
\eea

The dynamical equations \p{vectoreom4d} imply
\be
F^{(ij)} = \bar F^{(ij)} = 0~. \label{aux4d}
\ee
The lowest component of these superfields is a nonlinear analog
of the auxiliary field of the $N=2$ Maxwell theory.

Next, substituting the expressions \p{vectorih4d} for the spinor
Goldstone fermions in the l.h.s. of eqs. \p{components114d} and making use
of both \p{vectorchir4d} and \p{vectoreom4d}, we represent these l.h.s. as
\bea
&& \cD_{\alpha}^i\psi_{\beta}^j=
-\frac{i}{4}\left\{ \cD_{\alpha}^i,\cD_{\beta}^j\right\}W +
 \frac{i}{4}\varepsilon^{ij} \cD_{( \alpha}^k\cD_{\beta ) k}W  , \nn
&& \cDb_{\dot\alpha i}\psi_\alpha^j = -\frac{i}{2}
 \left\{ \cDb_{\dot\alpha i} , \cD_{\alpha}^j \right\} W \;.
\label{components24d}
\eea
Comparing \p{components24d}
with the definition \p{components114d} (taking into account \p{aux4d}),
it is straightforward to show that the objects $F^{(ij)}_{(\alpha\beta)}$,
$\bar F_{(ij)(\dot\alpha\dot\beta)}$, $X_{(ij)\alpha\dot\alpha}$
and $\bar X^{(ij)}_{\dot\alpha\alpha}$ satisfy a system
of homogeneous equations, such that the matrix of the coefficients
in them is nonsingular at the origin $W=\bar W = 0$. Thus these objects
vanish as a consequence of the basic equations:
\be
F^{(ij)}_{(\alpha\beta)} = \bar F_{(ij)(\dot\alpha\dot\beta)}
= X_{(ij)\alpha\dot\alpha}
= \bar X^{(ij)}_{\dot\alpha\alpha} = 0. \lb{vantens4d}
\ee

As a result, on shell we are left with the following superfield content:
\bea
&& \cD_{\alpha}^i\psi_{\beta}^j=
 \varepsilon^{ij}f_{\alpha\beta}\; , \quad
 \cDb_{\dot\alpha i}\psi_{j\alpha} =
\varepsilon_{ij} X_{\alpha\dot\alpha}\;, \nn
&& \cDb_{\dot\alpha i}{\bar \psi}_{\dot\beta j} =
 -\varepsilon_{ij}\bar f_{\dot\alpha \dot\beta}\;, \quad
 \cD_\alpha^i{\bar \psi}_{\dot\alpha}^j = \varepsilon^{ij}
   {\bar X}_{\dot\alpha\alpha} \;. \label{components14d}
\eea
The only new independent superfield at the dimension $(0)$ is the
complex one $F_{(\alpha \beta)}, \bar F_{(\dot\alpha\dot\beta)}$, while $A,
\bar A$ and $X_{\alpha\dot\beta}, \bar X_{\dot\alpha\beta}$ are
algebraically expressed through it and other independent superfields
as will be shown below. In the next Section
we shall show that this superfield is related, by an equivalence
field redefinition, to the Maxwell field strength obeying
the BI equation of motion.

Substituting the explicit expressions for the anticommutators
of covariant derivatives
into \p{components24d} and again using \p{components14d} in both sides
of \p{components24d},
we finally obtain:
\bea
&&
X_{\alpha\dot\alpha } = {\nabla}_{\alpha\dot\alpha}W+ \left(
X_{\gamma\dot\alpha}{\bar X}_{\dot\gamma\alpha} + \bar f_{\dot\alpha
\dot\gamma}f_{\alpha\gamma}\right)
{\nabla}^{\gamma\dot\gamma} W  ,\nn
&& A =
-{1\over 2} \bar X^{\;\;\beta}_{\dot\gamma}f_{\beta\gamma}
{\nabla}^{\gamma\dot\gamma}W~,
\label{sys14d}
\eea
It is easy to see that these algebraic equations indeed allow one to
express $A, \bar A$ and $X_{\alpha\dot\beta}, \bar X_{\alpha\dot\beta}$
in terms of $F_{(\alpha \beta)}$, $\bar F_{(\dot\alpha\dot\beta)}$ and
${\nabla}_{\alpha\dot\alpha}W $, ${\nabla}_{\alpha\dot\alpha}\bar W$:
\bea
&& X_{\alpha\dot\alpha} = {\nabla}_{\alpha\dot\alpha}W
+{1\over 2} ({\nabla}W\cdot {\nabla}W) {\nabla}_{\alpha\dot\alpha}\bar W
+ \ldots , \nn
&& A = {i\over 2}\, F^{(\beta}_{\;\;\gamma)}{\nabla}_{\beta\dot\gamma}\bar W
{\nabla}^{\gamma\dot\gamma}
W + \ldots~.
\eea

Returning to the issue of extracting an irreducible set of covariant
superfield projections of $W, \bar W$, it is easy to show that the further
successive action by covariant spinor derivatives on \p{components14d}
produces no new independent superfields. One obtains either
the equations of motion
(and Bianchi identities) for the independent basic superfields $W, \bar W,
\psi_{i\alpha}, \bar\psi^i_{\dot\alpha}$ and $F_{(\alpha\beta)}, \bar
F_{(\dot\alpha\dot\beta)}$, or some composite superfields
which are expressed through $x$-derivatives of the basic ones
(or as some appropriate nonlinear functions
of the basic superfields). The useful relations which essentially
simplify the analysis are the following ones:
\be
\left\{ \cD^{(i}_{\alpha}, \cDb^{j)}_{ {\dot\alpha}}\right\} =
\left\{ \cD^{(i}_{\alpha}, \cD^{j)}_{\beta}\right\} =
\left\{ \cDb^{(i}_{\dot\alpha}, \cDb^{j)}_{\dot\alpha}\right\}
= 0~. \lb{harmcov}
\ee
These relations are the covariant version of the
integrability
conditions for the Grassmann harmonic
$N=2$ analyticity \cite{GIKOS}. Thus the nonlinear $W, \bar W$
background specified by the equations \p{vectorih4d}-\p{vectoreom4d} respects
the Grassmann harmonic analyticity which plays a fundamental role
in $N=2, d=4$ theories.

Before going further, let us make a few comments.

First, the nonlinear realization setting we used,
in order to deduce our equations
\p{vectorih4d}-\p{vectoreom4d}, drastically differs from the standard
superspace differential-geometry setup of supersymmetric gauge theories
(see, e.g., \cite{bw}). The starting
point of the standard approach is the covariantization of the flat
derivatives (spinor and vector) by the gauge-algebra valued connections with
appropriate constraints on the relevant covariant superfield
strengths. In our case (quite analogously to the $N=1,
d=4$ and $N=1, d=3$ cases considered in Sect. 2 and 3) the covariant
derivatives include
no connection-type terms. Instead, they contain, in a highly non-linear
manner, the Goldstone bosonic $N=2$ superfields $W, \bar W$. These
quantities, after submitting them to the covariant constraints
\p{vectorih4d}-\p{vectoreom4d}, turn out to be the nonlinear-realization
counterparts of the $N=2$ Maxwell superfield strength. As we shall see,
the Bianchi identities needed to pass to the gauge potentials are
encoded in the set \p{vectorih4d}-\p{vectoreom4d}.

%In the differential-geometry approach the
%constraints like \p{vantens4d} emerge before going on shell,
%they are a consequence of the Bianchi identities.
In our nonlinear system we cannot separate in a simple way the kinematical
off-shell constraints from the dynamical on-shell ones. We could try to
relax our system by lifting the basic dynamical equations \p{vectoreom4d}
and retaining only the chirality condition \p{vectorchir4d} together
with \p{vectorih4d} and an appropriate covariantization of
the constraint (\ref{oshN2M4d}$b$). But in this case we
immediately face the same difficulty as in the
$N=2 \rightarrow N=1$ case:
a naive covariantization of (\ref{oshN2M4d}$b$) by replacing the flat
spinor derivatives by the
covariant ones proves to be not self-consistent. For self-consistency, it
should be properly modified order by order, without any clear guiding
principle. No such a problem
arises when the dynamical equations \p{vectoreom4d} are enforced. The
terms modifying the naive covariantization of (\ref{oshN2M4d}$b$)
can be shown to
vanish, as in the $N=2 \rightarrow N=1$ case \cite{BIK3}.

Nevertheless, there exists a highly nonlinear
field redefinition which relates the nonlinear superfield
Goldstone strength $W, \bar W$ to its flat counterpart
${\cal W}, \bar{\cal W}$
satisfying the off-shell irreducibility
conditions (\ref{oshN2M4d}). In this frame it becomes possible
to divide the kinematical and
dynamical aspects of our system and to write the appropriate off-shell
action giving rise to the dynamical equations, in a deep analogy
with the $N=2 \rightarrow N=1$ case \cite{BG2,RT}.

As the last comment, we note that all the fields of the multiplet comprised
by $W, \bar W$, except for $F_{(\alpha\beta)}, \bar
F_{(\dot\alpha\dot\beta)}$, can be given a clear interpretation as
Goldstone fields: $W \vert, \bar W \vert$ for the spontaneously broken
central-charge shifts, $\psi_{\alpha}^i\vert, \bar\psi_{i\dot\alpha}\vert$
for the spontaneously broken $S$-supersymmetry transformations and $F^{(ij)}
\vert, \bar F^{(ij)} \vert$ for the spontaneously broken $SO(5)/U(2)_{R}$
transformations.

\subsection{Bosonic equations of motion}
As the next important step in examining the superfield system
\p{vectorih4d}-\p{vectoreom4d}, we inspect its bosonic sector. The set of
bosonic equations can be obtained by acting on both sides of \p{components14d}
by two covariantized spinor derivatives, using the relations \p{sys14d} and
omitting the fermions in the final
expressions (which should contain only independent superfield projections
and their $x$-derivatives). Instead of analyzing the bosonic sector in full
generality, we specialize here to its two suggestive limits.

\noindent{\it 1. Vector fields limit.}
This limit amounts to
\be\label{1case4d}
 W \left|_{\theta=\bar\theta=0} =
 {\bar W}\right|_{\theta=\bar\theta=0} = 0 \; .
\ee
{}From eqs. \p{sys14d} with all fermions omitted, one can see that
\p{1case4d} imply
\be\label{1casea4d}
A={\bar A}=X_{\alpha\dot\alpha}={\bar X}_{\dot\alpha\alpha}=0 .
\ee
Thus, in this limit our superfields $W,{\bar W}$ contain only
$F_{\alpha\beta}, {\bar F}_{\dot\alpha \dot\beta}$ as the bosonic components,
which, owing to \p{vectoreom4d}, obey the following simple equations
\bea\label{case1eom4d}
\partial_{\alpha\dot\alpha} F^{\alpha\beta} -F_\alpha^\gamma
  {\bar F}_{\dot\alpha}^{\dot\gamma}\partial_{\gamma\dot\gamma}
F^{\alpha\beta}=0 \;\;\; \mbox{and c.c.}\;.
%&&\partial_{\alpha\dot\alpha} {\bar F}^{\dot\alpha \dot\beta}
%-F_\alpha^\gamma
%  {\bar F}_{\dot\alpha}^{\dot\gamma}\partial_{\gamma\dot\gamma}
%{\bar F}^{\dot\alpha \dot\beta}=0\;.
\eea
These equations coincide with eqs. \p{eom63d} of the $N=2 \rightarrow N=1$
case. So they can be handled in the same way, i.e. split into the
``true'' Bianchi identities and ``true'' equations of motion which
are the $d=4$ BI equations \p{last23d}.

Thus the superfield system \p{vectorih4d}-\p{vectoreom4d} encodes the
BI equation, in accord with the statement that this system provides
a supersymmetric extension of the latter.

\noindent{\it 2. Scalar fields limit.} This limit corresponds to
the reduction
\be\label{2case4d}
 \cD^i_{(\alpha} \cD_{\beta) i}W \left|_{\theta=\bar\theta=0} =
 \cDb^i_{(\dot\alpha}\cDb_{\dot\beta) i}
{\bar W}\right|_{\theta=\bar\theta=0} = 0 \; .
\ee
{}From eqs. \p{sys14d}  one finds that the reduction
conditions \p{2case4d} imply
\bea\label{2casea4d}
A=0,\; X_{\alpha\dot\alpha}=\partial_{\alpha\dot\alpha}W +
  X_{\gamma\dot\alpha}{\bar X}_{\dot\gamma\alpha}
\partial^{\gamma\dot\gamma}W
\eea
(and c.c.), while the equations of motion following from  \p{vectoreom4d} read
\bea\label{2caseb4d}
\partial_{\dot\alpha}^\alpha X_{\alpha\dot\beta} +
 {\bar X}^{\dot\gamma\alpha} X_{\;\;\dot\alpha}^\gamma
 \partial_{\gamma\dot\gamma} X_{\alpha\dot\beta}=0\;\;\;  \mbox{and c.c.}\;.
\eea
The system \p{2casea4d} can be easily solved
\bea\label{sol14d}
X_{\alpha\dot\alpha}=\partial_{\alpha\dot\alpha} W +
 \frac{\left( \partial W\right)^2 }{h}\partial_{\alpha\dot\alpha}
 {\bar W}\;\;\; \mbox{and c.c.}\,,
%&&{\bar X}_{\dot\alpha\alpha}=\partial_{\alpha\dot\alpha} {\bar W} +
\eea
where
\bea\label{sol1a4d}
&&h= 1- B +
 \sqrt{ \left( 1- B\right)^2 - C} , \nn
&& B= \left(\partial W\cdot
 \partial {\bar W}\right),\; C=\left( \partial W\right)^2
 \left( \partial{\bar W}\right)^2 .
\eea
One can check that the symmetric parts of eqs. \p{2caseb4d} are identically
satisfied with \p{sol14d} and \p{sol1a4d}. The trace part of \p{2caseb4d} can
be cast into the form: \bea\label{scalareq14d}
\partial_{\alpha\dot\alpha}\left( \frac{X^{\alpha\dot\alpha}+
 \frac{1}{2}X^2 {\bar X}^{\dot\alpha\alpha} }{1-\frac{1}{4}X^2{\bar X}{}^2}
 \right)=0 \;\;\; \mbox{and c.c.}\,.
\eea
Now, substituting \p{sol14d}, \p{sol1a4d} in \p{scalareq14d}, one finds
that the resulting form of these equations can be reproduced from the action
\be\label{NGaction4d}
S=\int
d^4x\left( \sqrt{ \left( 1-B\right)^2-C}-1\right) \;.
\ee
This action is the static-gauge form of the Dirac-Nambu-Goto
action of a 3-brane in D=6.

Thus, we have shown that the system
of our superfield equations \p{vectorih4d}-\p{vectoreom4d} is self-consistent
and gives a $N=2$ superextension of both the equations of $D=4$ BI theory
and those of the static-gauge 3-brane in $D=6$, with the nonlinearly
realized second $N=2$ supersymmetry. This justifies our claim that
\p{vectorih4d}-\p{vectoreom4d} are indeed a manifestly worldvolume
supersymmetric form of the equations of D3-brane in $D=6$ and,
simultaneously, of $N=2$ Born-Infeld theory. Similarly to the
previous examples, the nonlinear realization approach
yields the BI equations in a disguised form, with the Bianchi identities
and dynamical equations mixed in a tricky way. At the same time,
for the scalars we get the familiar static-gauge Nambu-Goto-type
equations. This is
in agreement with the fact that $W, \bar W$ undergo pure shifts
under the action of the central charge generators $Z, \bar Z$, suggesting
the interpretation of these superfields as the transverse brane coordinates
conjugated to $Z, \bar Z$. These generators, in turn, can be
interpreted as two extra components of the 6-momentum.

\subsection{Towards a formulation in terms of ${\cal W}, \bar{\cal W}$}

As was already mentioned, we expect that, like in the $N=1$
case \cite{BG2,RT}, there should exist an equivalence transformation
to a formulation
in terms of the conventional $N=2$ Maxwell superfield strength
${\cal W}, \bar{\cal W}$
defined by the off-shell constraints \p{oshN2M4d}.

A systematic, though as yet iterative procedure to
find such a field redefinition starts by passing to
the standard chirality conditions (\ref{oshN2M4d}$a$) from
the covariantly-chiral ones \p{vectorchir4d}. After some algebra,
\p{vectorchir4d} can be brought to the form
\be\label{chir24d}
{\bar D}{}_{\dot\alpha i} R=0 \;,
\quad D_{\alpha}^i {\bar R} =0 \;,
\ee
where
\bea\label{R4d}
R&=& W+\frac{1}{2}
 {\bar W}\left(\partial W\cdot \partial W\right) \nn
  &&+\frac{i}{4}D^\gamma_j W
  {\bar D}^{\dot\gamma j}{\bar W}\partial_{\gamma\dot\gamma}W  + O(W^5)\; .
\eea
Now we pass to the new superfields ${\cal W},{\bar{\cal W}}$ with preserving
the flat chirality
$$
\cW \equiv R\left(1-\frac{1}{2}{\bar D}{}^4{\bar R}{}^2\right)\quad
\mbox{and  c.c.}\;,  $$
\be
{\bar D}_{\dot\alpha i} \cW= D_\alpha^i \cWb =0 \;, \label{W4d}
\ee
where
\bea
D^4\equiv \frac{1}{48} D^{\alpha i}D^j_\alpha
D^{\beta}_{ i}D_{\beta j} \; , \quad \bar D^4 =
\overline{(D^4)} ~. \label{definitions14d}
\eea
Up to the considered third order, in terms of these superfields
eqs. \p{vectoreom4d} can be rewritten as
\bea
&& D^{ij} \cW = {\bar D}{}^{ij}\cWb\;, \label{bianchi4d} \\
&& D^{ij} \left( \cW + \cW\,{\bar D}{}^4{\cWb}^2 \right)  \nn
&& + \,{\bar D}{}^{ij} \left( \cWb +\cWb\, D^4{\cW}^2 \right) =0.\label{eom24d}
\eea
Eq. \p{bianchi4d} is recognized as the Bianchi identity (\ref{oshN2M4d}$b$), so
${\cal W}, \bar{\cal W}$ can be identified with the conventional
$N=2$ vector multiplet superfield strength. Eq. \p{eom24d} is then a
nonlinear generalization of the standard free $N=2$ vector multiplet
equation of motion \p{eqm14d}. The transformation properties of $\cW,\cWb$
can be easily restored from \p{unbrokentr4d}-\p{brokenZ4d} and the
definitions \p{R4d}, \p{W4d}.

The above procedure is an $N=2$ superfield analog of separating
Bianchi identities and dynamical equations for $F_{(\alpha\beta)},
\bar F_{(\dot\alpha\dot\beta)}$.
%In both
%cases we do not know the geometric principle behind the relevant field
%redefinitions.
Though in the bosonic case
we managed to find the appropriate field redefinition in a closed form,
we are not aware of it in the full superfield case. Nonetheless,
we can move a step further
and find the relation between
$W, \bar W$ and ${\cal W}, \bar{\cal W}$, as well as the
nonlinear dynamical equations for the latter,  up to the fifth order.
Then, using the transformation laws \p{unbrokentr4d}-\p{brokenZ4d}, we
can restore the hidden $S$-supersymmetry
and $Z, \bar Z$ transformations up to the fourth order.
In this approximation, the transformation laws and equations of motion read
\bea
&& \delta {\cal W} = f -{1\over 2}\bar D^4 (f \cAb_0 ) +
{1\over 4} \Box (\bar f \cA_0) \nn
&&+ {1\over 4i}
\,\bar D^{i\dot\alpha}\bar f D^{\alpha}_i\,\partial_{\alpha\dot\alpha}\cA_0~,
\quad \delta \bar{\cal W} = (\delta {\cal W})^*~,
\label{4transf4d} \\
&& \cA_0 = {\cal W}^2\left(1 + {1\over 2}\bar D^4 \bar{\cal W}^2
\right)~,\label{fdef4d} \\
&& f = c+ 2i\,\eta^{i\alpha}\theta_{i\alpha}~,
\\
&& D^{ij} B + \bar D^{ij} \bar B = 0~, \nn
&&B = {\cal W} +
{\cal W}\bar D^4 \left(\bar{\cal W}^2 + \bar{\cal W}^2D^4{\cal W}^2\right.\nn
&& + \left.
{1\over 2}\bar{\cal W}^2 \bar D^4 \bar{\cal W}^2 - {1\over 6} {\cal W}
\Box \bar {\cal W}^3\right)~. \lb{5eq4d}
\eea
The hidden supersymmetry transformations,
up to the third order, close on the $c, \bar c$ ones in the $\eta,
\epsilon$ and $\bar\eta, \bar\epsilon$ sectors, and on the standard $d=4$
translations in the $\eta,  \bar\eta$ sector. In the sectors $\eta, \eta$
and $\bar\eta,  \bar\eta$ the transformations commute, as it should be.
Note that \p{4transf4d} is already of the most general form
compatible with the chirality conditions and Bianchi identity \p{oshN2M4d}. So
this form will be retained to any order, only the functions $\cA_0$,
$\cAb_0$ will get additional contributions.

\section{OFF-SHELL ACTION OF THE N=2 BI THEORY}
\subsection{Embedding $N=2$ vector multiplet into a linear $N=4$ multiplet}
In the previous Section
we found the most general
hidden supersymmetry transformation law of ${\cal W}, \bar{\cal W}$
\p{4transf4d}
which is compatible with the defining constraints provided that the
$N=2$ superfield function $\cA_0$ is chiral
\be
\bar D_{\dot\alpha i} {\cal A}_0 =0 \;.
\label{chirA0}
\ee
By analogy with the $N=1$ construction of \cite{BG2}, in order to promote
\p{4transf4d} to a {\it linear} (though still inhomogeneous)
realization of the considered $N=4$ supersymmetry, it is natural
to treat ${\cal A}_0$ as a new {\it independent} $N=2$ superfield
constrained only by the chirality condition \p{chirA0} and to try to
define the transformation law of ${\cal A}_0$ under the $\eta, \bar \eta,
c, \bar c$-transformations in such a way that the
$N=2$ superfields ${\cal A}_0$,
${\cal W}, \bar{\cal W}$ form a closed set. Then, imposing a proper
covariant constraint on these superfields one could hope to recover
the structure \p{fdef4d} as the first terms in the  solution to this
constraint. In view of the covariance of this hypothetical constraint, the
correct transformation law for ${\cal A}_0$ to the appropriate order can be
reproduced by varying \p{fdef4d} according to the transformation law
\p{4transf4d}.
Since we know ${\cal A}_0$
up to the 4th order, we can uniquely restore its transformation law
up to the 3d order. We explicitly find
\be
\delta {\cA_0} = 2f\cW  +
{1\over 4} \bar f \Box \cA_1 + {1\over 4i}
\,\bar D^{i\dot\alpha}\bar f
D^{\alpha}_i\,\partial_{\alpha\dot\alpha} \cA_1,
\label{probe}
\ee
where
\be
{\cal A}_1 = {2\over 3}\,{\cal W}^3 + O({\cal W}^5)~,
\quad \bar D_{\dot\alpha i} {\cal A}_1 =0~.
\ee
We observe the appearance of a new composite chiral superfield ${\cal A}_1$,
and there is no way to avoid it in the transformation law \p{probe}. This is
the crucial difference from the $N=1$ case of ref. \cite{{BG2},{RT}} where
a similar reasoning led to a closed supermultiplet with only one extra
$N=1$ superfield besides the $N=1$ Goldstone-Maxwell one (the resulting
linear multiplet of $N=2$ supersymmetry is a $N=1$ superfield form
of the $N=2$ vector multiplet with a modified transformation
law \cite{part,ibmz}).

Thus, we are forced to incorporate a chiral superfield ${\cal A}_1$ as a new
independent $N=2$ superfield component of the linear $N=4$ supermultiplet we
are seeking. Inspecting the brackets of all these transformations
suggests that the only possibility to achieve their closure in
accord with the superalgebra \p{n4sa4d} is to introduce an {\it infinite}
sequence of chiral $N=2$ superfields and their antichiral conjugates
$\cA_n\;,~{\cAb}_n$, $n = 0, 1, \ldots$,
\be
\bar D_{\dot\alpha i} {\cA_n} =0~,\;
D_{\alpha}^i \bar{\cA_n} =0~, \label{defA}
\ee
with the following transformation laws:
\be
\delta {\cA_0} = 2f\cW  +
{1\over 4}\bar f \Box \cA_1 + {1\over 4i}
\,\bar D^{i\dot\alpha}\bar f D^{\alpha}_i\,
\partial_{\alpha\dot\alpha} \cA_1, \label{N4b1}
\ee
$$
\delta {\cA_1} = 2f \cA_{0}  +
{1\over 4}\bar f \Box \cA_{2} + {1\over 4i}
\,\bar D^{i\dot\alpha}\bar f D^{\alpha}_i\,\partial_{\alpha\dot\alpha}
\cA_{2} $$
\bea
&& ........... \nn
&& \delta {\cA_n} = 2f \cA_{n-1}  +
{1\over 4} \bar f \Box \cA_{n+1} \nn
&& + {1\over 4i}
\,\bar D^{i\dot\alpha}\bar f D^{\alpha}_i\,
\partial_{\alpha\dot\alpha} \cA_{n+1},\quad (n\geq 1)\label{N4b} \\
&& \delta \bar{\cA_n} = (\delta {\cA}_n)^*~ \nonumber
 \;.
\eea
It is a simple exercise to check that these transformations
close off shell both among
themselves and with those of the manifest $N=2$ supersymmetry
just according to the superalgebra \p{n4sa4d}.

Realizing (formally) the central charge generators
as derivatives in some extra complex ``central-charge coordinate'' $z$
\be
Z = {i\over 2} \frac{\partial}{\partial z}~, \quad \bar Z = {i\over 2}
\frac{\partial}{\partial \bar z}~, \label{defz}
\ee
and assuming all the
involved $N=2$ superfields to be defined on a $z, \bar z$ extension of the
standard $N=2$ superspace, it is instructive to rewrite the transformation
laws under the $c, \bar c$ transformations as follows:
\bea
\frac{\partial {\cal W}}{\partial z} =
\left( 1-\frac{1}{2}{\bar D}^4 \bar{\cal A}_0
\right)\;, \quad  \frac{\partial {\cal W}}{\partial\bar z}
=\frac{1}{4}\Box {\cal A}_0\;,
\label{cc1} \\
\frac{\partial {\cal A}_0}{\partial z} = 2{\cal W}\;, \quad \frac{\partial
{\cal A}_0}{\partial\bar z}= \frac{1}{4}\Box {\cal A}_1 \;, \label{cc2} \\
\frac{\partial
{\cal A}_n}{\partial z} = 2{\cal A}_{n-1}\;,
\quad \frac{\partial {\cal A}_n}{\partial\bar z}=
\frac{1}{4}\Box {\cal A}_{n+1} \;. \label{cc}
\eea
These relations imply, in particular,
\bea
&&\left(\frac{\partial^2}{\partial z\partial {\bar
z}} -\frac{1}{2} \Box \right) {\cal W}=0\; , \nn
&&\left(\frac{\partial^2}{\partial z\partial {\bar z}} -\frac{1}{2} \Box
\right) {\cal A}_n=0~. \label{mass}
\eea
If we regard $z, \bar z$ as the actual coordinates,
which extend the $d=4$ Minkowski
space to the $D=6$ one, the relations \p{mass} mean that the constructed
linear supermultiplet is on shell from the $D=6$ perspective. On the other
hand, from the $d=4$ point of view this multiplet is off-shell, and the
relations \p{cc1} - \p{cc}, \p{mass} simply give
a specific realization of the central
charge generators $Z, \bar Z$ on its $N=2$ superfield components. In this
sense this multiplet is similar to the previously known special $N=2, d=4$
and $N=4, d=4$ supermultiplets, which are obtained from the on-shell
multiplets in higher dimensions via non-trivial dimension reductions and
inherit the higher-dimensional translation generators as non-trivially
realized central charges in $d=4$ \cite{FS,SSW}.
Since the superalgebra \p{n4sa4d} is just a $d=4$ form of the $N=(2,0)$ (or
$N=(0,2)$) $D=6$ Poincar\'e superalgebra, it is natural to think that
the above supermultiplet has a $D=6$
origin and to try to reveal it. For the time being we prefer to treat the
above
 infinite-dimensional
representation in the pure $d=4$ framework
as a linear realization of the partial
spontaneous breaking of the central-charge extended $N=4, d=4$
supersymmetry \p{n4sa4d} to the standard $N=2$ supersymmetry.

\subsection{Invariant action}
In the approach proceeding
from a linear realization of PBGS,
the Goldstone superfield Lagrange density is, as a rule,
a component of the same linear supermultiplet to which
the relevant Goldstone
superfield belongs. This is also true for the case under consideration.
A good candidate for the chiral $N=2$ Lagrangian density
is the superfield $\cA_0$.
Indeed, the ``action''
\be
S=\int d^4x d^4\theta \cA_0 + \int d^4x d^4\bar\theta \cAb_0 \label{action1}
\ee
is invariant with respect to the transformation \p{N4b1} up to surface
terms.
With the interpretation of the central charge
transformations as shifts with respect to the coordinates $z, \bar z$, the
action \p{action1} does not depend on these coordinates in virtue of eqs.
\p{cc2}, though the Lagrangian density can bear such a dependence.

It remains to define covariant constraints
which would express $\cA_0$,  $\cAb_0$ in terms
of ${\cal W}$, $\bar{\cal W}$, with preserving the linear representation
structure \p{4transf4d}, \p{N4b1}, \p{N4b}. Because an infinite number
of $N=2$ superfields ${\cal A}_n$ is present in our case,
there should exist an infinite set of constraints trading
all these superfields
for the basic Goldstone ones ${\cal W}$, $\bar{\cal W}$.

As a first step in finding these constraints let us note that
the following expression:
\bea\label{constr1}
\phi_0 &=& \cA_0 \left( 1-\frac{1}{2}{\bar D}{}^4\cAb_0\right) -
\cW^2 \nn
&&- \sum_{k=1}\frac{ (-1)^k}{2\cdot 8^k}
   \cA_k\Box^k {\bar D}{}^4 \cAb_k
\eea
is invariant, with respect to the $f$ part of
the transformations \p{4transf4d}, \p{N4b1}, \p{N4b}.
This leads us to choose
\be\label{constr1a}
\phi_0=0
\ee
as our first constraint. For consistency with $N=4$ supersymmetry,
the constraint \p{constr1a} should be invariant with respect
to the full transformations \p{4transf4d}, \p{N4b1}, \p{N4b}, with the
$\bar f$ part taken into account as well. We shall firstly specialize
to the $\bar c$ part of the $\bar f$ transformations.
The requirement of the $\bar c$ covariance produces
the new constraint
\bea\label{constr2}
&&\phi_1=\Box \cA_1 +2\left( \cA_0\Box \cW - \cW\Box\cA_0 \right)\nn
&&-\sum_{k=0}\frac{ (-1)^k}{2\cdot8^k}
\left( \Box \cA_{k+1}\Box^k{\bar D}{}^4 \cAb_k \right. \nn
&& -\left.
  \cA_{k+1} \Box^{k+1} {\bar D}{}^4 \cAb_k    \right) =0\;.
\eea
It is invariant under the $f$ transformations, but requiring it
to be invariant also under the $\bar c$ part gives rise
to the new constraint
\bea\label{constr3}
\phi_2 =\Box^2 \cA_2 +2\left( \cA_0\Box^2 \cA_0
- \Box\cA_0\Box\cA_0 \right.\nn
\left. +2\Box\cA_1\Box\cW
   -\cA_1\Box^2\cW -\cW\Box^2\cA_1 \right) \nn
-\sum_{k=0}\frac{ (-1)^k}{2\cdot8^k}
\left( \Box^2 \cA_{k+2}\Box^k{\bar D}{}^4 \cAb_k \right.\nn
-\left.
     2\Box \cA_{k+2}\Box^{k+1}{\bar D}{}^4 \cAb_k +
    \cA_{k+2} \Box^{k+2} {\bar D}{}^4 \cAb_k    \right) =0\nonumber
\eea
and so on. The full infinite set of constraints is by construction
invariant under the $f$ and $\bar c$ transformations. Indeed,
using the relations \p{cc1}-\p{cc} one may
explicitly check that
\be
\frac{\partial \phi_n}{\partial z} = 0~, \quad
\frac{\partial \phi_n}{\partial \bar z} = {1\over 4}\phi_{n+1}~,
\ee
so the full set of constraints is indeed closed.

The variation of the basic constraints with respect to
the $\bar f$ transformations has the following general form:
\be
\delta \phi_n = \bar\eta{}^{i\dot\alpha}{\bar\theta}_{i\dot\alpha}
{\cal B}_n + \bar\eta{}^{i\dot\alpha} ({\cal F}_n)_{i\dot\alpha} \;.
\ee
Demanding this variation to vanish gives rise to the two sets of constraints
\be\label{cccc}
(\mbox{a})\;\; {\cal B}_n=0 \;, \quad (\mbox{b})\;\;
({\cal F}_n)_{i\dot\alpha} =0 \;.
\ee
The constraints (\ref{cccc}a) are easily recognized as those
obtained above from
the $\bar c$ covariance reasoning. One can show by explicit
calculations that
\be
{\bar D}{}^{i\dot\alpha} ({\cal F}_n)_{j\dot\beta} \sim \delta^i_j
    \delta^{\dot\alpha}_{\dot\beta} {\cal B}_n \;.
\ee
Thus the fermionic constraints (\ref{cccc}b) seem to be more
fundamental.
In order to prove that the basic fermionic constraints
(\ref{cccc}b) are actually equivalent to the
bosonic ones (\ref{cccc}a), one has to know the
general solution to {\it all} constraints.
For the time being we have explicitly checked this
important property only for the iteration solution
given below. Taking for granted that this is true in general,
we can limit our
attention to the type $(\mbox{a})$ constraints only. The constraints
\p{constr1},
\p{constr2} are just of this type.

At present we have no idea, how to explicitly solve
the above infinite set of constraints and find a closed
expression for the Lagrangian densities ${\cal A}_0$, $\bar{\cal A}_0$
similar to the one known in the $N=2 \rightarrow N=1$ case \cite{BG2}.
What we are actually able to do, so far, is to
restore a general solution by iterations. E.g., in order to restore
the action up to the 8th order, we have to know
the following orders in $\cA_k$:
\bea
 \cA_0= \cW^2 + \cA_0^{(4)}+\cA_0^{(6)}+\cA_0^{(8)}+\ldots \; ,\nn
 \cA_1= \cA_1^{(3)}+\cA_1^{(5)}+\cA_1^{(7)}+\ldots \; , \nn
 \cA_2= \cA_2^{(4)}+\cA_2^{(6)}+\ldots \;, \quad \cA_3
= \cA_3^{(5)}+\ldots \; .
\eea
These terms were found to have the following explicit structure:
\bea\label{An}
&&\cA_0^{(4)}= \frac{1}{2}\cW^2 {\bar D}{}^4\cWb^2\;, \nn
&&  \cA_0^{(6)}=\frac{1}{4}{\bar D}{}^4
\left[ \cW^2\cWb^2\left( D^4\cW^2+{\bar D}{}^4\cWb^2\right)\right.\nn
&& -\left.
   \frac{1}{9}\cW^3\Box\cWb^3\right], \nn
&&\cA_0^{(8)} =\frac{1}{8}{\bar D}{}^4\left[ 4\cW^2\cAb_0^{(6)}
+ 4\cWb^2\cA_0^{(6)} \right. \nn
&&+
  \cW^2\cWb^2D^4\cW^2{\bar D}{}^4\cWb^2
-\frac{2}{9}\cW^3\Box\left(\cWb^3D^4\cW^2\right)  \nn
&& \left. -\frac{2}{9}\cW^3{\bar D}{}^4\cWb^2\Box\cWb^3
+\frac{1}{144}\cW^4\Box^2\cWb^4 \right] \;, \nn
&&\cA_1^{(3)} =  \frac{2}{3}\cW^3\;, \quad \cA_1^{(5)}
=\frac{2}{3}\cW^3{\bar D}{}^4\cWb^2 \;, \nn
&&\cA_1^{(7)}= {\bar D}{}^4
\left[ \frac{1}{2}\cW^3\cWb^2{\bar D}{}^4\cWb^2+
   \frac{1}{3}\cW^3\cWb^2D^4\cW^2 \right. \nn
&& -\left. \frac{1}{24}\cW^4\Box\cWb^3 \right] \;, \nn
&&\cA_2^{(4)} =  \frac{1}{3}\cW^4\;,\quad \cA_2^{(6)}
=  \frac{1}{2}\cW^4{\bar D}{}^4\cWb^2\;,\nn
&&  \cA_3^{(5)}=\frac{2}{15}\cW^5 \;.
\eea

Note that, despite the presence of growing powers
of the operator $\Box$ in our constraints,
in each case the maximal power of $\Box$ can be finally taken off
from all the terms in the given
constraint, leaving us with this maximal power of $\Box$ acting on
an expression which starts from the appropriate ${\cal A}_n$.
Equating these final expressions
to zero allows us to algebraically express all ${\cal A}_n$
in terms of ${\cal W}, \bar{\cal W}$
and derivatives of the latter. For example,
for $\cA_3^{(5)}$ we finally get the following equation:
\be
\Box^3 \cA_3^{(5)} =\frac{2}{15} \Box^3 \cW^5 \; \Rightarrow \;
\cA_3^{(5)} =\frac{2}{15} \cW^5~.
\ee

This procedure of taking off the degrees of $\Box $ with discarding
possible ``zero modes'' can be justified as follows: we are interested
in an off-shell solution that preserves the manifest standard
$N=2$ supersymmetry including the Poincar\'e covariance.
This rules out possible on-shell
zero modes as well as the presence of explicit $\theta$'s or $x$'s
in the expressions which remain after taking off
the appropriate powers of $\Box$.
It can be checked to any desirable order that these ``reduced''
constraints yield correct local expressions
for the composite superfields $\cA_n$,
which prove to transform just in accordance with the original
transformation rules
\p{4transf4d}, \p{N4b1}, \p{N4b}. We have explicitly verified
this for our iteration
solution \p{An}.

The explicit expression for the action, up to the 8th order in
${\cal W}, \bar{\cal W}$, reads
\bea
&&S^{(8)} = \left(\int d^4xd^4\theta \cW^2+\mbox{c.c.}\right)\nn
&&+
\int dZ \left\{ \cW^2\cWb^2\left[
 1+\frac{1}{2}\left( D^4\cW^2 +{\bar D}{}^4\cWb^2\right)\right] \right. \nn
&& -\frac{1}{18}\cW^3\Box\cWb^3
+\frac{1}{4} \cW^2\cWb^2\left[ \left( D^4\cW^2
+{\bar D}{}^4\cWb^2\right)^2 \right. \nn
&& \left.+ D^4\cW^2{\bar D}{}^4\cWb^2
  \right] -\frac{1}{12}D^4\cW^2\cWb^3\Box\cW^3 \nn
&&\left. -\frac{1}{12}{\bar D}{}^4\cWb^2\cW^3\Box\cWb^3
+\frac{1}{576}\cW^4 \Box^2 \cWb^4 \right\} \;. \label{8}
\eea
This action, up to a slight difference in the notation, coincides with the
action found by Kuzenko and Theisen \cite{KT} from the requirements of
self-duality and invariance under nonlinear shifts of ${\cal W}, \bar{\cal
W}$ (the $c, \bar c$ transformations in our notation). Let us point out that
the structure of nonlinearities in the $c, \bar c$ transformations of
${\cal W}, \bar{\cal W}$ in our approach is uniquely fixed
by the original $N=4$
supersymmetry transformations and the constraints imposed. 
The next, 10th
order part of the $N=4$ invariant $N=2$ BI action
 can be easily restored
from eqs. \p{An}. Its explicit form looks not too enlightening,
so  we do not present it here. Actually, the action can be restored in this
way to any order. It is of interest to find it in a closed form (if existing).

Finally, let us point out that after doing the $\theta $ integral,
the pure Maxwell field strength part of the
bosonic sector of the above action (and of the hypothetical
complete action) comes entirely from the expansion
of the standard Born-Infeld bosonic action. Just in this sense the
above action is a particular $N=2$ extension of the bosonic BI action.
The difference from the action of ref. \cite{Ket} is just in
higher-derivative
terms with the $\Box$ operators.
These correction terms are crucial for the invariance
under the hidden $N=2$ supersymmetry, and they
drastically change, as compared to ref. \cite{Ket}, the structure
of the bosonic action, both in the pure scalar fields sector and the
mixed sector involving couplings
between the Maxwell field strength and the
scalar fields. By a reasoning
of \cite{BIK3}, the additional terms are just those needed
for the existence of an equivalence field
redefinition bringing the scalar fields action into the
standard static-gauge Nambu-Goto form.

\section{N=4 BI THEORY}
Finally, we derive the superfield equations of $N=4$ BI
theory.

The $N=4, D=4$ Maxwell theory \cite{n4ym} is described
by the covariant strength superfield
${\cal W}_{ij}=-{\cal W}_{ji}, (i,j=1,\ldots,4)$,
satisfying the following independent
constraints \cite{Sohn}
\be
\bar{\cal  W}^{ij} \equiv \left( {\cal W}_{ij}\right)^* =
\frac{1}{2}\varepsilon^{ijkl}{\cal W}_{kl}\;,\label{realW}
\ee
\be D_{\alpha}^k {\cal W}_{ij}-\frac{1}{3}
\left(\delta_i^k D_{\alpha}^m {\cal W}_{mj} -
   \delta_j^k D_{\alpha}^m {\cal W}_{mi} \right) =0.
\label{n8eomflat}
\ee
In contrast to the $N=2$ gauge theory, no off-shell
superfield formulation exists in the $N=4$ case: the constraints
\p{realW}, \p{n8eomflat} put the theory on shell.

As in the $N=2$ case, in order to construct a nonlinear generalization
of \p{realW}, \p{n8eomflat}
one should firstly define the appropriate algebraic framework. It is given
by the following central charge-extended $N=8, D=4$ Poincar\'e superalgebra:
\bea
 \left\{ Q_{\alpha}^i, {\bar Q}_{\dot{\alpha}j}
        \right\}=2\delta^i_jP_{\alpha\dot{\alpha}}, \;
\left\{ S_{\alpha}^i, {\bar S}_{\dot{\alpha}j}
        \right\}=2\delta^i_jP_{\alpha\dot{\alpha}}, \nn
\left\{ Q_{\alpha}^i,  S_{\beta}^j
   \right\}=\varepsilon_{\alpha\beta} Z^{ij}, \;
      \left\{ {\bar Q}_{\dot{\alpha}i}, {\bar S}_{\dot{\beta}j}
        \right\}=
 \varepsilon_{\dot{\alpha}\dot{\beta}}{\bar Z}_{ij},
 \label{n8sa}
\eea
\be
{\bar Z}_{ij}=\left(Z^{ij}\right)^* =
\frac{1}{2}\varepsilon_{ijkl}Z^{kl}.\label{reality}
\ee
This is a $D=4$ notation for the type IIB Poincar\'e superalgebra
in $D=10$.

We wish the $N=4, D=4$ supersymmetry $\left\{ P_{\alpha\dot{\alpha}},
 Q_{\alpha}^i, {\bar Q}_{\dot{\alpha}j}\right\}$ to remain unbroken,
so we are
led to introduce the Goldstone superfields
\bea\label{gfn8}
&&Z^{ij}\Rightarrow W_{ij}(x,\theta,\bar\theta)\;, \quad
S_{\alpha}^i\Rightarrow \psi^{\alpha}_i (x,\theta,\bar\theta)\;,\nn
&&{\bar S}_{\dot{\alpha}j}\Rightarrow
{\bar\psi}_i^{\dot\alpha}(x,\theta,\bar\theta)\;.
\eea
The reality property \p{reality} automatically implies the
constraint \p{realW} for $W_{ij}$:
\be
\bar W^{ij} = \frac{1}{2}\varepsilon^{ijkl} W_{kl}~. \lb{real1}
\ee

On the coset element $g$
\bea\label{n8coset}
g&= &\mbox{exp}\,i\left(-x^{\alpha\dot{\alpha}}P_{\alpha\dot{\alpha}}
   +\theta^{\alpha}_i Q^i_{\alpha}+{\bar\theta}^i_{\dot\alpha}
    {\bar Q}_i^{\dot\alpha}\right)\nn
 && \mbox{exp}\,i\left(\psi^{\alpha}_i S^i_{\alpha}+{\bar\psi}^i_{\dot\alpha}
    {\bar S}_i^{\dot\alpha} + W_{ij}Z^{ij} \right)
\eea
one can realize the entire $N=8, D=4$ supersymmetry \p{n8sa} by left
shifts. The Cartan forms (except for the central charge one) and covariant
derivatives formally coincide with \p{vectorcf4d}-\p{fullcd4d},  the
indices $\left\{ i,j\right\}$ now ranging from 1 to 4. The central charge
Cartan form reads:
\bea\label{n8ccform}
\omega^Z_{ij}&=&dW_{ij}+\frac{1}{2}\left(
d\theta^{\alpha}_i\psi_{\alpha j}-
 d\theta^{\alpha}_j\psi_{\alpha i}\right.\nn
&&+\left.
\varepsilon_{ijkl}d{\bar\theta}^k_{\dot\alpha}
    {\bar\psi}{}^{\dot\alpha l}\right)~. \;
\eea
By construction, it is covariant under all transformations of the
$N=8, D=4$  Poincar\'{e} supergroup. The Goldstone superfields
$\psi_{\alpha i}$ and
${\bar\psi}^k_{\dot\alpha}$ can be covariantly eliminated by the
inverse Higgs procedure,
as in the previous case. The proper constraint reads as follows:
\be\label{n8con1}
\omega^Z_{ij} \vert_{d\theta, d\bar\theta} =0 \;.
\ee
It amounts to the following set of equations:
\bea
&&(a)\;\cD_{\alpha}^k W_{ij}+\frac{i}{2}\left( \delta_i^k \psi_{\alpha j}-
     \delta_j^k\psi_{\alpha i} \right) =0\;, \nn
&&(b) \; \cDb_k^{\dot\alpha}W_{ij}+\frac{i}{2}
\varepsilon_{ijkl}{\bar \psi}^{\dot\alpha l}
   =0 \;,  \label{eq2}
\eea
which are actually conjugated to each other in virtue of \p{real1}. We
observe that, besides expressing the fermionic Goldstone superfields through
the basic bosonic one $W_{ij}$:
\be\label{eq3} \psi_{\alpha
i}=-\frac{2i}{3}\cD_{\alpha}^jW_{ij}\;, \quad {\bar \psi}^{\dot\alpha
i}=-\frac{2i}{3}\cDb_j^{\dot\alpha}{\bar W}^{ij} \;,
\ee
eqs. \p{eq2} impose the nonlinear constraint
\be\label{n8eom}
\cD_{\alpha}^k W_{ij}-\frac{1}{3}\left(\delta_i^k \cD_{\alpha}^m W_{mj} -
   \delta_j^k\cD_{\alpha}^m W_{mi} \right) =0
\ee
(and its conjugate). This is the sought nonlinear generalization
of \p{n8eomflat}.

It is straightforward to show that eq. \p{n8eom} implies the disguised form
of the BI equation  for the nonlinear analog of the abelian
gauge field strength. For the six physical bosonic fields $W_{ij}\vert$
we expect the equations corresponding to the static-gauge of
$3$-brane in $D=10$ to hold. Thus eqs. \p{real1}, \p{n8eom} plausibly
give a manifestly worldvolume
supersymmetric description of D3-brane in a flat $D=10$ Minkowski background.
No simple off-shell action can be constructed in this case, since
such an action
is unknown even for the free $N=4$ Maxwell theory. But even the construction
of the physical fields component action for this $N=4$ BI theory is of
considerable interest. We hope to study this system in more detail
elsewhere.

\section{CONCLUSIONS}
We presented a systematic approach to deducing the
dynamics of superbranes from nonlinear realizations of the appropriate PBGS
patterns in a way manifestly covariant under the  worldvolume supersymmetry.
In the case of D-branes the corresponding equations simultaneously describe
the appropriate superextensions of BI theory. We also proposed a procedure
for constructing the off-shell superfield action of $N=2, d=4$ BI theory
associated  with the PBGS pattern $N=4 \rightarrow N=2$ (D3-brane in $D=6$).

Among the problems for further study let us distinguish generalizing the above
consideration to non-abelian super BI theories and incorporating non-trivial
curved backgrounds into the PBGS framework. As a first important step in
tackling the second task it would be tempting to find PBGS formulations of
superbranes with the $AdS_n \times S^m$ bosonic parts. As a more technical
problem, we finally mention finding out the precise correspondence of the
PBGS examples considered here with the superembedding approach to superbranes
\cite{Dima} along the lines of refs. \cite{ACGHN}, \cite{PST}, \cite{BPPST}.
In \cite{PST}, \cite{BPPST} such a correspondence has been
established  for the $N=1, D=4$ supermembrane and space-filling D3-brane.

\section*{Acknowledgements}

We are grateful to Eric Bergshoeff,
Sergio Ferrara, Sergei Ketov, Sergei Kuzenko,
Dmitri Sorokin and Mario Tonin for useful discussions. This work was
supported in part by the Fondo Affari Internazionali Convenzione Particellare
INFN-JINR, grants RFBR-CNRS 98-02-22034, RFBR-DFG-99-02-04022,
RFBR 99-02-18417 and NATO Grant PST.CLG 974874.

\end{document}